\documentclass[twocolumn,preprintnumbers,amsmath,amssymb]{revtex4}
\usepackage{graphicx,epsfig}
\usepackage{dcolumn}
\usepackage{bm}
\begin{document}

\title{Anisotropic Local Stress and Particle Hopping in a Deeply 
Supercooled Liquid}

\author {Sarika Bhattacharyya and Biman Bagchi\footnote[1]
{For correspondence: bbagchi@sscu.iisc.ernet.in}}
\affiliation{ Solid State and Structural Chemistry Unit,
Indian Institute of Science,
Bangalore 560012, India}

\begin{abstract}
The origin of the microscopic motions that lead to 
stress relaxation in deeply supercooled liquid remains unclear.
We show that in such a liquid the stress relaxation is 
locally anisotropic which can serve as the driving force 
for the hopping of the system on its free energy surface.
However, not all hopping are equally effective in relaxing the
local stress, suggesting that diffusion can decouple from
viscosity even at local level. On the other hand, orientational relaxation
is found to be always coupled to stress relaxation.
\end{abstract} 

\maketitle
Dynamics of supercooled liquid show many fascinating properties,
namely the non-exponentiality in various relaxation functions of dynamical
variables, 
like the stress, density, composition, polarization and 
orientation\cite{sastry, 
angell,arnab,parisi,lepo}.
One often observes a very slow  power law decay in the intermediate 
to long times\cite{fayerprl}.
Although these have drawn attention of experimentalists and theoreticians 
and have been extensively studied, many aspects still remain ill-understood.
It is believed that the nature of the microscopic motion undergoes a 
drastic change 
at a temperature T$_{B}$ (a temperature substantially below 
the melting temperature, T$_{m}$). At the temperature $T_{B}$, the
continuous hydrodynamic type motion, which is prevalent at higher temperature  
changes to discontinuous motion. It is also believed that T$_{B}$ is close to 
the temperature where effects of the underlying free energy landscape on the 
dynamics are felt for the first time\cite{sastry}.
$T_{B}$ is also found to be close  to the mode coupling 
temperature, T$_{c}$. 
This change in the nature of the microscopic motion is 
believed to be the origin of the 
experimentally observed $\alpha$-$\beta$ bifurcation \cite{jg} 
and also the cross-over between the rotational and translational relaxation 
times \cite{ediger}. 

Due to the complexity of the problem, computer simulations have played a key
role in augmenting our understanding in this area. In particular, simulations
allow one to directly look at the microscopic events.
The computer simulation studies of the stress auto correlation function 
in the supercooled liquid could successfully reproduce the 
power law behavior of the stress auto-correlation function 
(SAF) \cite{arnab,parisi}.
However, in the deeply supercooled liquid, one finds that within the
simulation time, the relaxation, 
after an initial decay (typically 10-20$\%$) becomes fully arrested. 
The microscopic origin of the subsequent decay 
is unclear.
The computer simulation studies further show that
the orientational and translational 
hopping of particles are the only mode present and hence the stress 
relaxation can happen only via hopping. 
However, since the 
relaxation time is much much longer (could be of the order of ms or sec), 
the computer simulation results, 
(which can explore mostly upto nano second regime and sometimes micro second 
regime), cannot
include the effects of these hopping. Therefore, one cannot explore
the relationship between hopping and the {\em total} stress relaxation.
  
The experiments on the other hand are successful in showing the 
decay of the SAF. However, except a recent work using
single molecule spectroscopy \cite{van}, these experiments are macroscopic
and do not provide enough information of the microscopic 
motions in the system. 

Majumdar \cite{majum} had  earlier discussed the importance of 
local relaxation modes (of wavelength less than the short range order)
in giving rise to non-exponentiality 
in the stress relaxation function. This work discussed relaxation in
terms of relaxation within small regions, surfaces and also volumes,
with progressive lengthening of time scales.
However, in that analysis the
basic mechanism of relaxation was still assumed to be continuous.

In this Letter  we demonstrate for the first time that 
in the deeply supercooled liquid (where hopping is the only 
surviving large amplitude motion), there is
a close relationship between the {\em local} stress
and the 
orientational and translational hopping. 
 The local SAF is anisotropic and is found to change drastically 
during 
the hopping, thus showing that the local stress and the hopping of 
a particle are intimately connected. The anisotropy in the local stress 
could be the driving force for hopping. 
As the free energy of the
system can be expressed in terms of the position dependent stress in a
generalized Ginzburg-Landau formulation \cite{landau}, 
the change of the anisotropic
stress due to hopping should be regarded as the driving force for the 
transitions of the system between different minima of the free energy surface.  
However, not all hoppings are effective in relaxing the stress. 

 Our solvent is represented by binary Lennard-Jones mixture, which has been 
extensively studied \cite{sastry, sastry1, kob1,others} and is known 
to be a good glass former,
and our  solute probes are prolate ellipsoids.  
Pressure is kept constant by Andersen's piston method \cite{andersen} 
while in the case of temperature, a damped oscillator method has been 
adopted which keeps temperature constant at 
each and every time step \cite{brown}. The piston mass involved here is 
$0.0027(m_{A}/\sigma_{A}^4)$ which is regarded as optimum \cite{brown,haile}.
The interaction between two 
ellipsoids with arbitrary orientations is assumed to be given by the
Gay-Berne (GB) potential \cite{gb},
\begin{eqnarray}
U_{GB}& = &4\epsilon(\hat r,\hat u_1,\hat u_2)\Biggl[ \left( \frac{\sigma_{0}}
{r-\sigma(\hat r,\hat u_1,\hat u_2)+ \sigma_{0}} \right) ^{12}\\ \nonumber
&&-\left (\frac{\sigma_{0}}{r-\sigma(\hat r,\hat u_1,\hat u_2)+ \sigma_{0}}\right )^{6} \Biggr]
\end{eqnarray}
where $\hat u_1$ $\hat u_{2}$ are the axial vectors of the ellipsoids 1 and 2.
 $\hat r$ is the 
vector along 
the intermolecular vector $r = r_2 - r_1$, where $r_1$ and $r_2$ denote the 
centers of mass of ellipsoids 1 and 2 respectively. $\sigma(\hat r,\hat u_1,
\hat u_2)$ and 
$\epsilon(\hat r,\hat u_1,\hat u_2)$ are the orientation-dependent range and 
strength parameters
respectively. $\sigma$ and $\epsilon$ depend on the aspect ratio $\kappa$.
The minor axis of the ellipsoid is equal to the diameter of the larger 
solvent and the major axis is 3 times that of the minor axis.
Finally, the interaction between a sphere and an ellipsoid is accounted for by 
a modified  GB-LJ potential given below
\begin{equation}      
	U_{Ei} = 4\epsilon_{Ei}\left [ \left(\frac{\sigma(\theta)_{Ei}}{r} 
\right )^{12} - \left (\frac{\sigma(\theta)_{Ei}}{r}\right )^6\right ]
\end{equation}
\noindent where 'E' denotes the ellipsoids and 'i' can be 'A' or 'B'. 
The expression for $\sigma(\theta)_{Ei}$ is available \cite{sbor}.

The ellipsoid in binary mixture system with the above mentioned potential is a
well behaved system and it can also exhibit the experimentally observed 
anomalous viscosity dependence of the orientational correlation time 
\cite{sbor}. Four ellipsoids were placed far from each other in a binary 
mixture of 500 particles with number of 'A' particles, $N_{A}=400$ and number
of 'B' type particles $N_{B}=100$. 
The reduced temperature is expressed as,
$T^{*}$(=$k_{B} T/\epsilon_{A}$), the reduced pressure as, 
$P^{*}$(= $P\sigma^{3}_{A}/\epsilon_{AA}$).
and the reduced density as $\rho^{*}$(=$\rho \sigma_{A}^{3}$). 
The time is scaled by $\tau =\sqrt{(m_{A}\sigma_{A}^{2}/\epsilon_{AA})}$.  
The time step of the simulation is .002 $\tau$ and the system 
is equilibrated for 1.5 $\times$ 10$^{5}$ steps and the data collection step 
is 5 $\times$ 10$^{6}$. The studies have been performed at T$^{*}$=0.8 and the 
P$^{*}$=6 and 10. 

At P$^{*}$ =6, both hopping and continuous motions exist in the system, thus 
the stress relaxation could not be directly correlated with the hopping.
At P$^{*}$=10, only microscopic motion that survives is hopping. 
In a recent study,  
we have reported observation of correlated translational and 
orientational hopping \cite{sab} at this pressure.
After extensive simulations we could find only two different kinds of motions.
The translational hopping was either associated with 
correlated hopping of 5-6 nearest neighbors or it exhibited a  
motion in a ring like tunnel. While it is possible that other types of 
motion like isolated hopping can exist, we could not find them. 
The hopping rate was found to be $2\times 10^{7}$ where both the type of 
motions occurred with almost equal frequency.
In the following we
focus on the stress relaxation and its relation vis-a-vis hopping 
at P$^{*}$=10 and T$^{*}$=0.8. The reduced density of the system is 1.41.

  In figure 1 we show two different kinds of spatial hopping observed in our 
simulations. Both the hoppings are associated with orientational 
hopping. In the inset we also plot the orientational time correlation 
functions (OCF), before, during and after the hopping. 
 Figure 1(a) shows the trajectory of the first tagged 
ellipsoid and inset shows its orientational time correlation function, 
$<P_{2}(\hat u_{i}(0)\hat u_{i}(t))>/{<P_{2}(\hat u_{i}(0)\hat u_{i}(0))>}$,
were $P_{2}$ is the second order Legendre polynomial.
The hopping takes place in 20$\tau$ and the displacement is 0.5$\sigma$. 
Here the ellipsoid hopping is accompanied by hopping of 5-6 of its nearest 
neighbors. The OCF decays only during the period of hopping. 
In figure 1 (b) the trajectory of the second tagged ellipsoid 
and in the inset its orientational time correlation functions are shown. 
Note that in this case the displacement of the particle is large 
(1.1$\sigma$) and it also takes place over a very long period of time
(50 $\tau$). Here we find that 
the tagged particle is moving in a ring like tunnel. Although the 
orientational correlation function decays during the hopping, 
it's decay is less when compared to that of the 1st tagged particle.

\begin{figure}
\epsfig{file=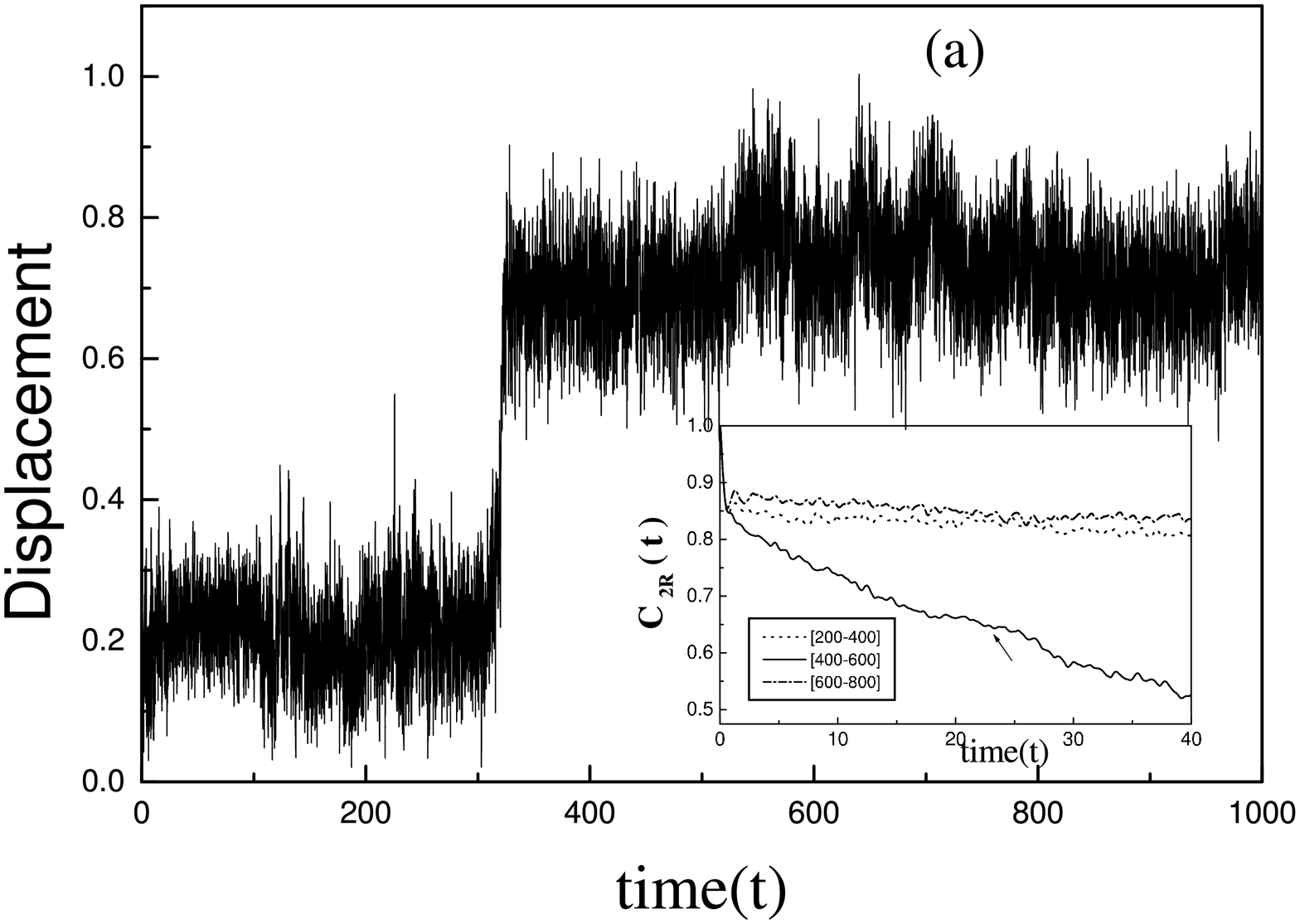,height=7cm,width=6cm,angle=-90}
\epsfig{file=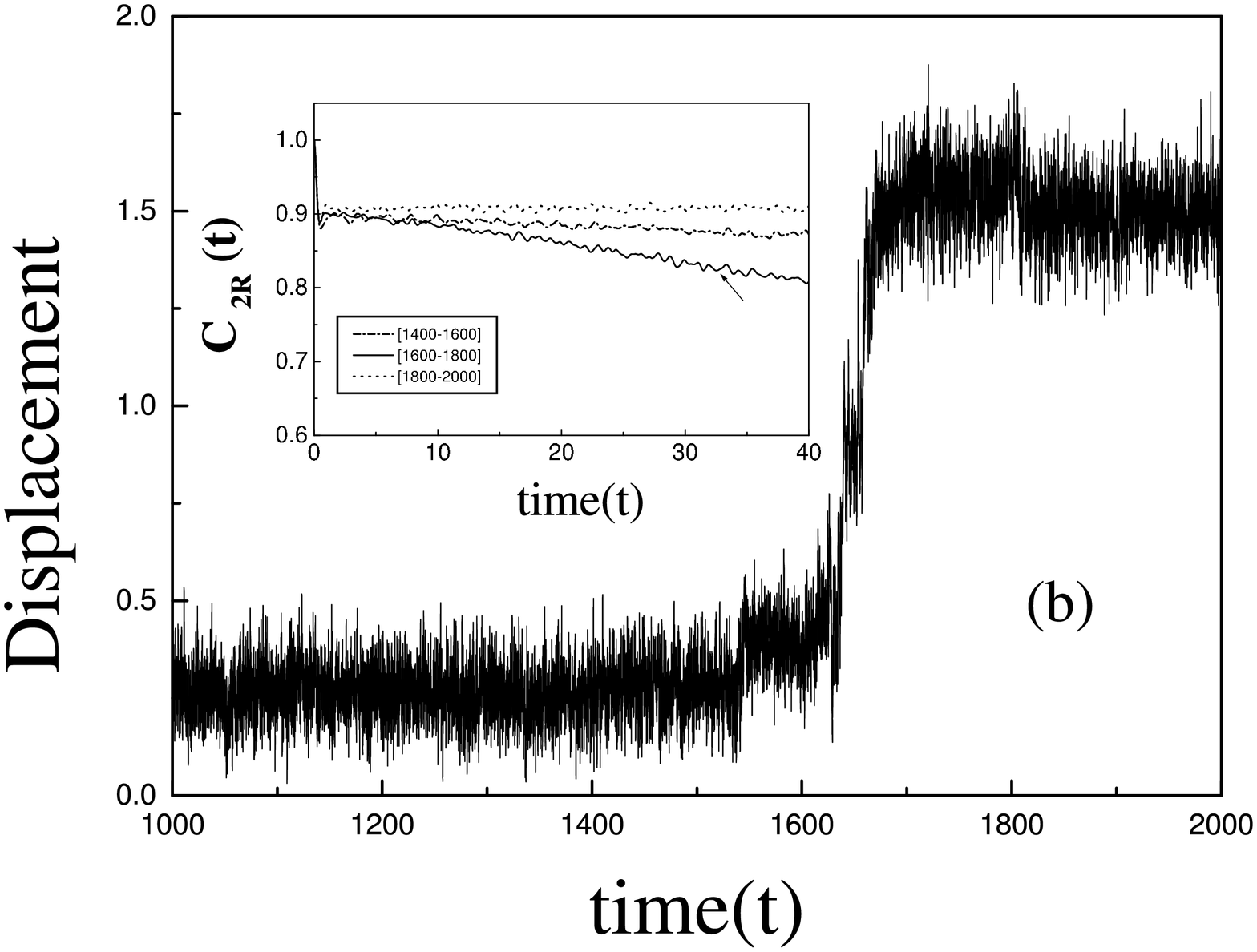,height=7cm,width=6cm,angle=-90}
\caption{ (a) Shows the displacement of the 1st tagged ellipsoid over 
1000 $\tau$. There is a hopping of the ellipsoid around 300 $\tau$. In the 
inset we plot the orientational correlation function obtained at different 
intervals. (b) Same as figure 1 (a) but for the 2nd tagged ellipsoid.
The plots are at P$^{*}$=10 and T$^{*}$=0.8.}
\end{figure}

The motions in a string like and ring like tunnel in a deeply supercooled 
liquid have been 
reported earlier by other authors, although they did not discuss this large 
displacements over a long time for the ring like motion \cite{sharon}. 

In the following part we discuss in details the local stress 
relaxation associated with these two different kinds of hoppings.
The local stress around the ellipsoid is obtained by summing over the 
stress on the ellipsoid and its nearest neighbors.
The stress has six components and the stress auto-correlation function 
is given by $<\sigma_{ij}(0)\sigma_{ij}(t)>$, where i,j=x,y,z.
In case of only Lennard-Jones fluid, $\sigma_{ij}=\sigma_{ji}$, but 
for particles interacting via GB and GB-LJ potential this is not so.

Figures 2 (a) and (b) show the SAF around the 
1st tagged ellipsoid, before and after the period it is hopping,
respectively. 
Before the hopping there is an anisotropy of the stress.
The xy and yx components 
of the stress are much larger than the others 
and also their correlations cease to decay. 
{\it This anisotropy leads to a 
hopping of the ellipsoid, mostly in the z direction}. During the hopping 
there is a relaxation of the SAF in xy and yx components and after hopping 
all the components relax. Note that the total stress in all the components 
are also lower.

\begin{figure}
\epsfig{file=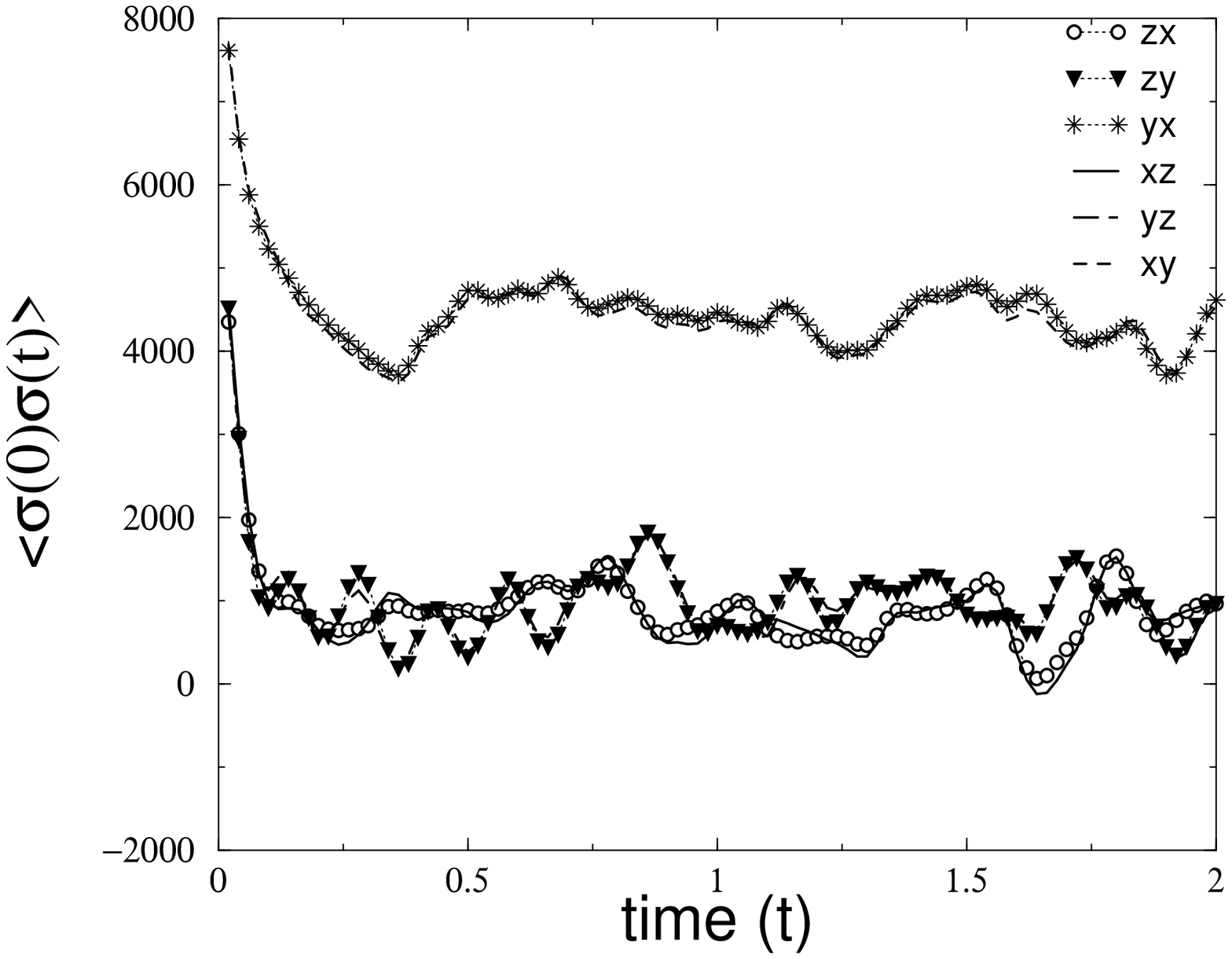,height=5cm,width=8cm}
\hspace*{-0.0cm}
\epsfig{file=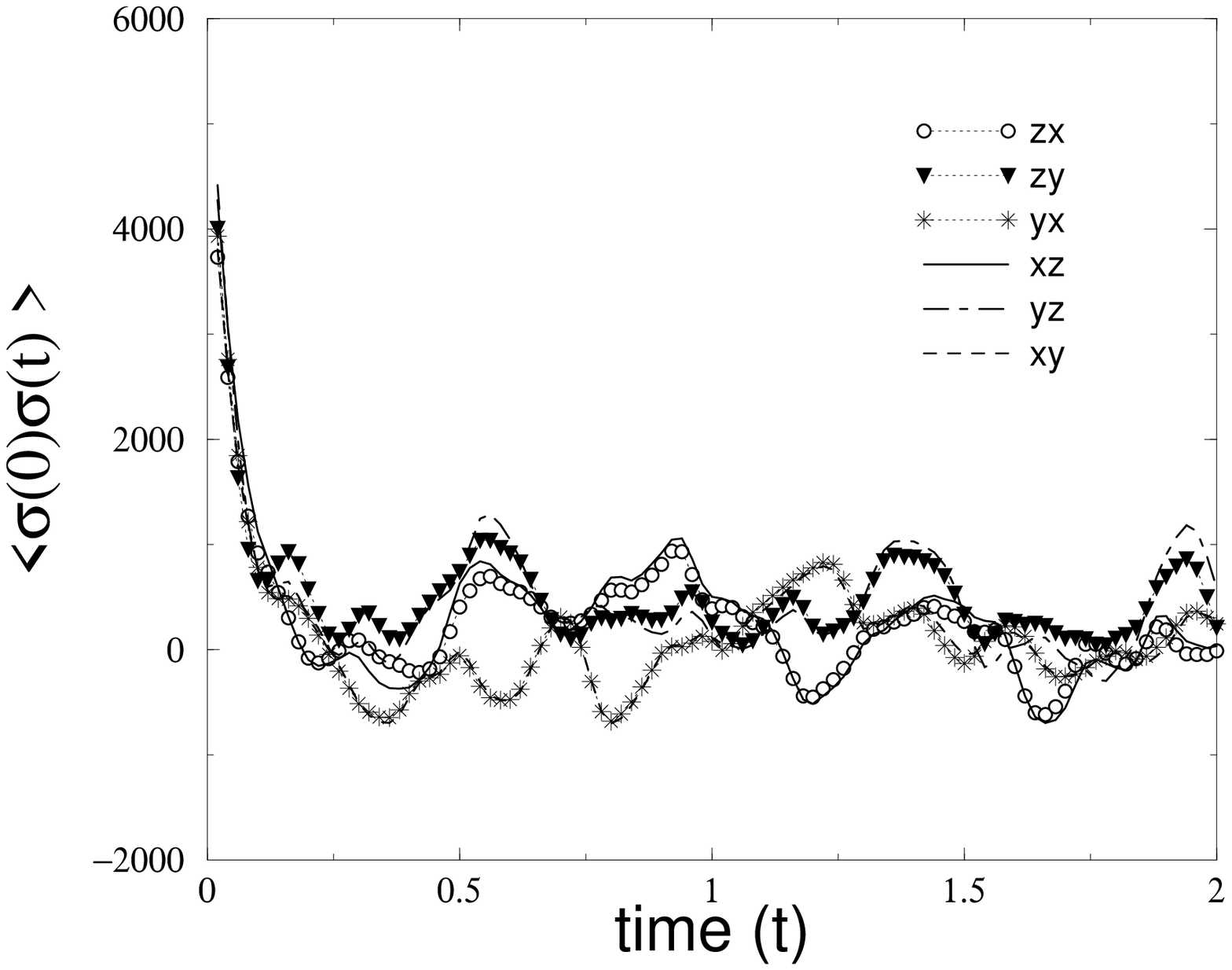,height=5cm,width=8cm,angle=0}
\caption{(a) The different components of the local stress 
auto-correlation (obtained from the sum of the 
stress of the 1st ellipsoid and its nearest neighbors)
function obtained before hopping, between 320-330 $\tau$. 
(b)is the same as 2 (a) but the 
components of the SAF are obtained after hopping between (340-350$\tau$).
The plots are at P$^{*}$=10 and T$^{*}$=0.8.}
\end{figure}

Figures 3 (a) and (b) show the SAF around the 
2nd tagged ellipsoid, before and after the period it is hopping, 
respectively. 
Before the hopping there is an anisotropy of the stress. The yz and zy 
components 
of the stress are much larger than others, and 
also their correlations cease to decay. {\it This leads to the hopping of 
the ellipsoid, mostly 
in the x direction}.
During and after the hopping there is an exchange of the stress. After the 
hopping although the yz component of the stress relaxes and also the t=0 
value of all the components reduces,
the SAF in the xz and zx components cease to decay. 
Thus, this kind of motion in a ring like 
tunnel does not lead to the relaxation of all the 
components of the stress.
\begin{figure}
\epsfig{file=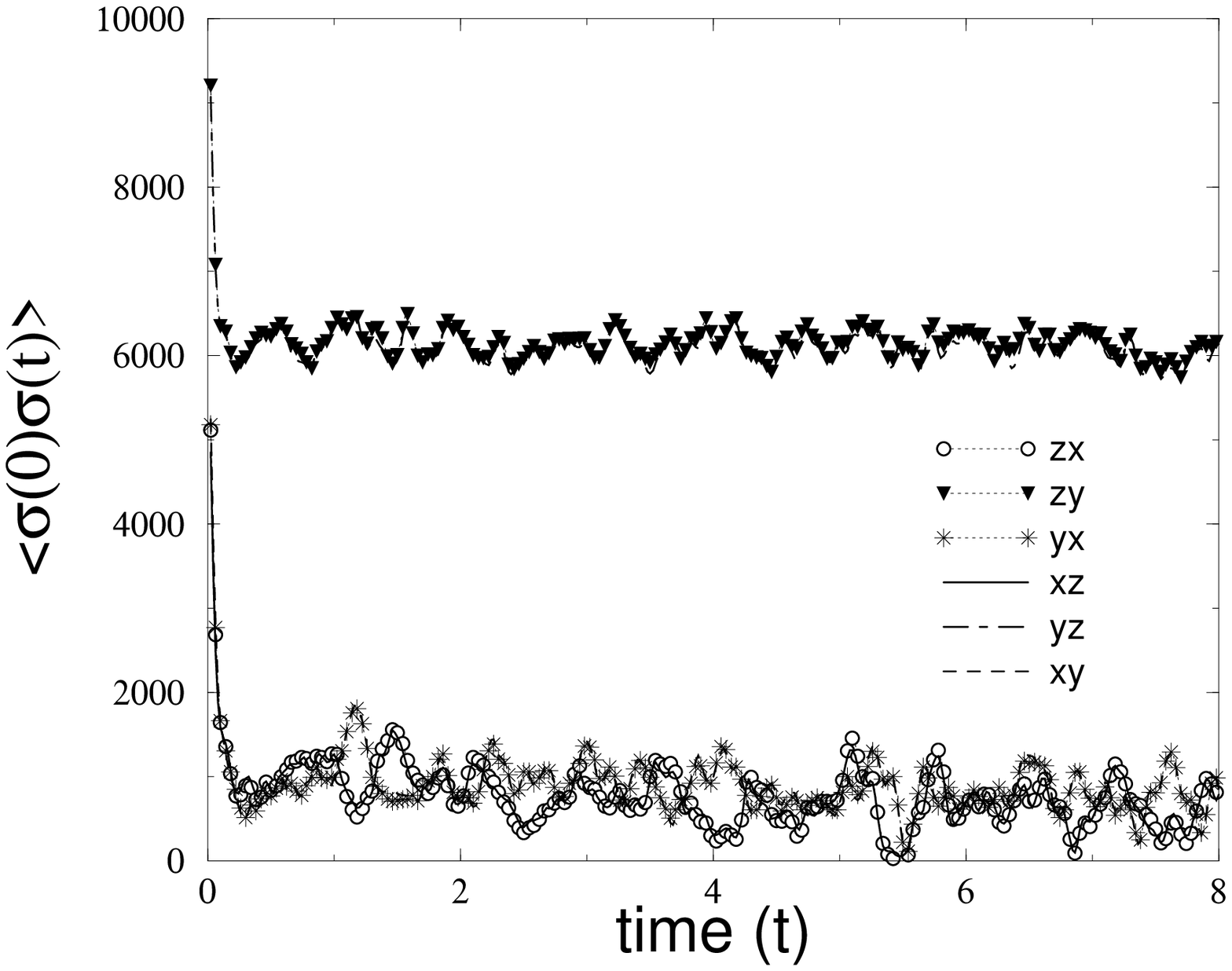,height=5cm,width=8cm,angle=0}
\vspace*{-0.0cm}
\hspace*{-0.0cm}
\epsfig{file=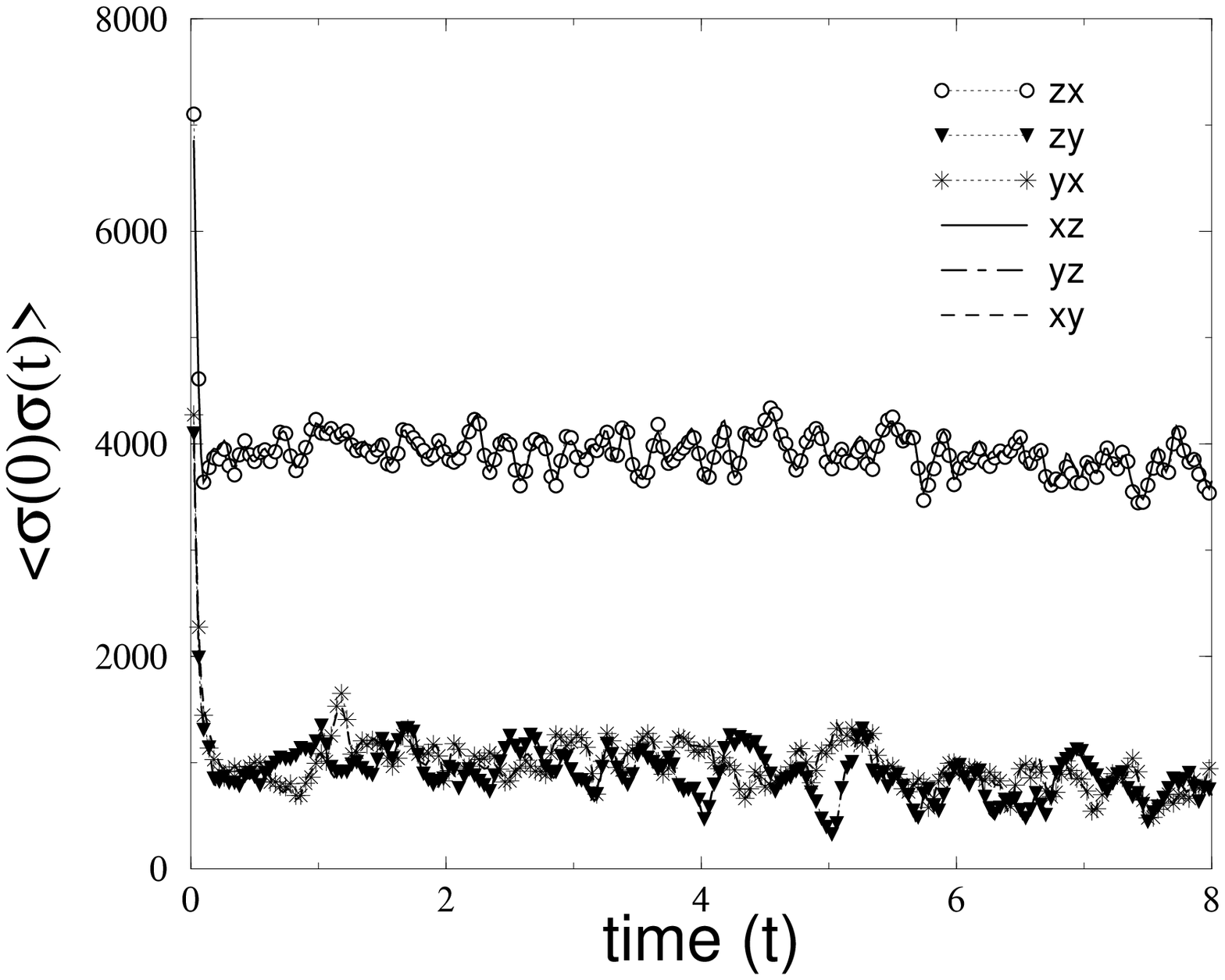,height=5cm,width=8cm,angle=0}
\caption{ (a)The different components of the local stress 
auto-correlation (obtained from the sum of the 
stress of the 2nd ellipsoid and its nearest neighbors)
function obtained before hopping, between 1590-1630 $\tau$. 
(b) is the same as 3 (a) but the 
components of the SAF are obtained after the hopping,   
between 1670-1710$\tau$.
The plots are at P$^{*}$=10 and T$^{*}$=0.8.}
\end{figure}

The orientational correlation function of the 1st tagged particle 
relaxes more ( inset of figure 1 (a)) compared to that of the 2nd tagged 
particle (inset of figure 1(b)) when computed in the respective 
intervals where they are hopping.
From figure 2 and 3 we found that the local stress relaxation takes place 
when the 1st ellipsoid is hopping where as when the second ellipsoid is 
hopping although there is an exchange of stress between it's components, the local 
SAF does not completely relax. There is a direct connection between the 
local stress and the orientational relaxation functions, implying that 
rotation and viscosity are coupled even in a localized region.

In order to understand what happens to the surrounding of the local
 region when there is a relaxation of stress due to hopping we have 
studied the stress 
auto correlation function of a bigger region of 2 $\sigma$ around the 
1st ellipsoid.We found there are about 62-67 particles 
in this region. Although there is an anisotropy of the 
components of the stress in this larger region, this anisotropy cannot 
be correlated with the 
direction of hopping. Some of the components of SAF build up immediately 
after hopping and in a later time it relaxes.
Similar analysis when done in a even bigger region shows that it takes 
longer for the stress of that region to relax and also the effect of the 
hopping is less.

In conclusion we demonstrated  that
 the {\it direction} of the hopping of the tagged particle is 
determined by the anisotropy in the stress.  
 Anisotropic stress relaxation is different when there is a many-particle 
hopping and there is a motion in a ring like tunnel.
 Although there is an exchange of stress between the components due to the 
particle motion, the stress relaxation is less in a ring like motion.
Interestingly,  the effect of hopping is found to spread over the adjoining region like 
ripples with 
the amplitude decreasing with increasing distance from the ellipsoid.
We note that in the case of the second tagged ellipsoid (Fig.1b) 
although it 
translates more, the stress relaxation during its hopping is less. Thus 
suggesting that translational motion and viscosity are decoupled even in a
localized region. On the other hand, the orientational relaxation and also 
the stress relaxation is more 
for the first ellipsoid.
Thus suggesting that the orientational motion always remains
coupled to viscosity. 
This is in agreement with the experimental results and in fact provides a
microscopic explanation of the results which are known for a long time.
There can be an apparent connection between the stress tensor and the 
momentum circulation. Thus it is possible that the non-decaying SAF implies 
that momentum circulation exists in a deeply supercooled liquid 
 Since the anisotropic stress contributes to the free energy of the system, 
a change in the anisotropy drives the system from one free energy minimum to 
the other. When the anisotropy in the stress disappears and all the SAF 
relaxes then the system definitely moves to a lower free energy minimum.

 This work was supported by a grant from CSIR, India. We thank Srikanth Sastry 
Arnab Mukherjee and Rajesh Murarka for helpful discussions.

\end{document}